\RequirePackage{fix-cm}

\documentclass[5p,sort&compress]{elsarticle}

\usepackage[utf8]{inputenc}
\usepackage{textcomp}
\usepackage{graphicx}
\usepackage{amsmath,amssymb}
\usepackage{tabularx,booktabs}
\usepackage{gensymb}
\usepackage{siunitx}
\usepackage[dvipsnames]{xcolor}
\usepackage{microtype}
\usepackage{balance}
\usepackage{xurl}
\usepackage[hidelinks]{hyperref}

\DeclareUnicodeCharacter{2032}{\ensuremath{^\prime}}
\DeclareSIUnit{\atomicpercent}{at.\%}
\DeclareSIUnit{\wtpercent}{wt.\%}
\DeclareSIUnit{\angstrom}{\text{\AA}}
\DeclareSIUnit{\h}{\hour}
\date{}
\hypersetup{
  pdftitle={Suppression of Grain Boundary Nitrogen Degradation by Boron in Superalloys},
  pdfauthor={Cameron Crabb; Geri Topore; James O. Douglas; Mark C. Hardy; Michele S. Conroy; David Dye}
}

\begin{document}
\begin{frontmatter}
\title{Suppression of Grain Boundary Nitrogen Degradation by Boron in Superalloys}

\author[IC]{Cameron Crabb\corref{cor1}}\ead{c.crabb23@imperial.ac.uk}
\author[IC]{Geri Topore}
\author[IC]{James O. Douglas}
\author[RR]{Mark C. Hardy}
\author[IC]{Michele S. Conroy\corref{cor1}}\ead{m.conroy@imperial.ac.uk}
\author[IC]{David Dye\corref{cor1}}\ead{d.dye@imperial.ac.uk}

\address[IC]{Department of Materials, Royal School of Mines, Imperial College London, Prince Consort Road, London, SW7 2BP, UK}
\address[RR]{Rolls-Royce plc, Elton Road, Derby, DE24 8BJ, UK}

\cortext[cor1]{Corresponding authors.}

\begin{abstract}
Superalloys enable the hottest sections of jet engines to operate at high temperatures, but their lifetime is limited by environmentally assisted degradation. In polycrystalline alloys, oxygen consumption by the surface scale allows nitrogen to penetrate beneath the surface oxide, creating brittle internal nitrides along grain boundaries that may promote cavities, voids and cracks from hot static or cyclic deformation. Boron is added in trace amounts to improve grain boundary performance, yet whether it merely strengthens existing boundaries or actively controls oxidation-induced reaction products remains unclear. Here we show that boron suppresses grain boundary nitrogen degradation in a Co/Ni-base $\gamma/\gamma^{\prime}$ strengthened superalloy by segregating to the interfaces of growing internal nitrides. Both alloys develop stratified chromia/alumina scales at 800°C in 100 h, but the B-free variant produces larger, deeper and more elongated Ti/Nb-rich grain boundary nitrides. Correlative 4D-Scanning Transmission Electron Microscopy (4D-STEM) and Atom Probe Tomography (APT) identify these particles as chemically complex B1 nitrides and reveal boron enrichment at the nitride--matrix interface. Density-corrected Gibbsian excess analysis confirms boron segregation to the boundary, showing that trace boron limits nitride growth by controlling the nitride--matrix interface.
\end{abstract}


\end{frontmatter}

\section{Introduction}
Ni-base superalloys are used in high temperature applications where performance is frequently governed not only by yield strength and creep, but also by the ability of grain boundaries to resist environmentally assisted cracking \cite{Reed2006TheApplications,Pollock2006Nickel-BasedProperties,Coakley2010CoarseningSuperalloy,Coakley2011CreepSuperalloy,JKnowles2019APrecipitates}. Previous studies have found that a certain amount of boron is required in superalloys to provide resistance to oxidation-assisted grain boundary cracking, e.g. during fatigue in service, but that too high a boron content also results in degradation \cite{Kontis2016OnSuperalloy,Xiao2004Effect718}. With the advent of fine grain powder metallurgy alloys in the 1970s, the amount of boron required had to be reformulated. It has since become evident that grain boundary borides (e.g. M$_5$B$_3$, where M will often be largely Mo or Cr) oxidise easily ahead of crack tips, providing a rationale for the maximum limit \cite{Kontis2016OnSuperalloy,Kitaguchi2024MC900C}. It has long been thought that boron segregates to grain boundaries, but the mechanism by which boron protects the material against oxidation-assisted cracking has remained mysterious, an example of the emerging topic of grain boundary design \cite{Kontis2017OnSuperalloy,Kontis2016OnSuperalloy}.

Polycrystalline Ni-base superalloys are used in rotor discs and also in static components such as casings, rings and bolts, where oxidation resistance, creep resistance and grain boundary damage tolerance must be maintained simultaneously near the upper service window of current wrought alloys \cite{Pollock2006Nickel-BasedProperties,Reed2006TheApplications,Hardy2020SolvingSuperalloys,Hardy2020DevelopingRotors}. There is a wide variety of processing routes used depending on the application, with cast and wrought routes for many static components, powder metallurgy used in fatigue-sensitive rotating parts, single crystals used in the hottest turbine blades, fusion welding for joining components and additive manufacturing used to reduce part count or increase supply chain resilience in small volume, intricate parts \cite{Reed2006TheApplications,Donachie2002Superalloys:Guide,Pollock2006Nickel-BasedProperties,Caron1999EvolutionApplications,Panwisawas2020MetalSuperalloys,Hunziker2000OnWelding}. 

As increased operating temperatures result in improved efficiency and lower emissions, the environmental degradation challenge has become increasingly demanding. In polycrystalline alloys, oxygen and nitrogen grain boundary diffusion leads to subsurface precipitation and grain boundary embrittlement \cite{Krupp2000SelectiveSuperalloys,Pfaendtner2001Oxygen-inducedEmbrittlement,Kitaguchi2013OxidationSuperalloy}. Following the discovery of the L1$_{2}$-ordered $\text{Co}_3(\text{Al},\text{W})$ phase by Sato and co-workers, Co-base and later Co/Ni-base $\gamma/\gamma'$ superalloys emerged as promising candidates for high-temperature applications \cite{Sato2006Cobalt-baseAlloys,SUZUKI2008High-temperatureAlloys,Pollock2010NewAlloys,Vorontsov2022PrecipitateCreep,Yan2014Micromechanics}. The primary benefits of the high-Co Ni superalloys appear to be in polycrystalline applications where the reduced freezing range results in improved castability, weldability and printability, which along with the depressed $\gamma'$ solvus results in enhanced forgeability. Early Co-base $\gamma/\gamma'$ alloys nevertheless showed poor oxidation resistance because continuous, adherent protective scales were difficult to establish \cite{Klein2011HighSuperalloys,Klein2013TheAlloys}. Progressive incorporation of Ni and Cr into these chemistries improved oxidation behaviour and led to Co/Ni-base alloy candidates that challenge conventional wrought superalloys \cite{Klein2013TheAlloys,Neumeier2015NovelResistance,Murray2020APrinting,Tang2021Alloys-by-design:Manufacturing}.

At intermediate temperatures, the external scale develops through selective oxidation, in which elements with the highest oxygen affinity and sufficient mobility form protective or semi-protective oxides. In Ni- and Co/Ni-base superalloys, this commonly produces chromia-rich scales, with alumina becoming increasingly important where alloy chemistry and exposure conditions favour the development of a more stable Al-rich sublayer \cite{Giggins1971Oxidation1200C,Klein2011HighSuperalloys,Yan2014AlloyingAlloys,Prescott1992TheAlloys,Ismail2017AlloyingSuperalloys}. However, the formation and regeneration of a protective external scale is a necessary but not necessarily sufficient condition to resist environmental degradation. Even when a continuous scale is present, reactive species can penetrate along grain boundaries, where local oxygen potential, nitrogen activity and solute availability differ substantially from the bulk alloy. These subsurface reactions are particularly significant because they occur at the same grain boundaries that control intergranular fracture, creep damage and tensile ductility \cite{Nemeth2017Environmentally-assistedSuperalloy,Kitaguchi2013OxidationSuperalloy}. Strain to fracture is therefore an important measure of the mechanical performance of oxide scales in polycrystalline superalloys, since cracking or fracture of the scale can compromise its protective function.

Among these subsurface reactions, internal nitridation is often observed but has been studied much less than external oxide-scale growth. Protective external scale formation reduces the oxygen potential, allowing nitrogen to penetrate further into the alloy, particularly along fast grain boundary diffusion paths. In alloys containing strong nitride-forming elements such as Ti, Al and Nb, this then promotes the formation of internal nitrides. Previous work has most commonly reported these features as Ti- and Al-rich nitrides, while the possible role of Nb-rich or mixed Ti/Nb nitrides remains far less resolved \cite{Cruchley2013ChromiaTitanium,Krupp1999InternalSystem,Krupp1999InternalDescription,Smith2005TheAlloys}. This reflects a broader limitation in the current understanding of internal nitridation: many studies identify nitride formation chemically, but comparatively little structural or interfacial analysis has been performed to determine the crystallography, local solute redistribution and interface chemistry of these phases. Nitrides can be especially damaging under slow strain-rate conditions in air, where environmental ingress and stress-assisted grain boundary damage can act together, leading to severe reductions in ductility and component lifetime \cite{Nemeth2017Environmentally-assistedSuperalloy,Pfaendtner2001Oxygen-inducedEmbrittlement,Kitaguchi2013OxidationSuperalloy,Texier2016CrackInclusions}.

The chemistry of modern Co/Ni-base superalloys further complicates matters. As well as Cr and Al to form chromia and alumina, and the strengthening Ni$_3$Al $\gamma'$ phase, additions of Ti and Ta are made to increase stacking fault energy and hence precipitate strength, W and Mo to promote matrix strength and reduce diffusion-driven coarsening, together with C to form MC for grain boundary pinning and castability. There is also evidence that nil C alloys suffer from enhanced grain boundary oxidation \cite{Cruchley2013ChromiaTitanium,Reed2006TheApplications,Jena1984TheSuperalloys,Shen2019The600}. Ti and Nb are particularly significant because they are strong nitride-forming elements and are frequently implicated in internal grain boundary reaction products in oxidised superalloys \cite{Krupp1999InternalSystem,Cruchley2013ChromiaTitanium,Sanviemvongsak2020IntergranularManufacturing}. Mo, W and Ta also contribute to boride and carbide precipitation, including tetragonal M$_5$B$_3$ and M$_3$B$_2$ borides, cubic (B1) MC carbides and complex cubic M$_6$C carbides \cite{Kontis2016OnSuperalloy,Reed2006TheApplications,Jena1984TheSuperalloys}. 

The effects of boron have been the subject of detailed investigation. It has been held to improve creep resistance by segregating to grain boundaries, where it has been speculated that it might reduce boundary energy, strengthen cohesion and/or suppress grain boundary sliding or cavitation at elevated temperature \cite{Kontis2016OnSuperalloy,Letellier1994DirectTomography,Blavette1996Atomic-scaleSuperalloys}. Boron may also influence the redistribution of solutes along grain boundaries during oxidation. We have recently shown that boron can redistribute to alumina-rich fingers within the oxide scale, highlighting that boron participates in local reaction-front chemistry \cite{Crabb2026TimeDependentSuperalloy}. Despite this, its role in oxygen-assisted grain boundary cracking and oxidation-induced ductility loss remains poorly understood. In particular, it is not known whether boron simply segregates to pre-existing grain boundaries, or whether it actively modifies nitride/matrix interfaces and suppresses the growth of damaging internal nitrides.

Internal oxidation and nitridation are common degradation pathways in Ni- and Co-base superalloys and emerging compositionally complex or high-entropy alloys containing strong nitride- and oxide-formers such as Ti, Al, Nb, Ta and Cr \cite{Kitaguchi2013OxidationSuperalloy,Esmaily2020High-temperatureAlloys,Butler2015High-TemperatureAlloys}. This raises the broader alloy design question: can trace interfacial solutes be used deliberately to control the chemistry, mobility and morphology of internal oxidation or nitridation fronts? If so, minor additions such as boron should not be viewed only as conventional grain boundary strengtheners, but as active interfacial dopants.

Resolving such behaviour experimentally is challenging. Scanning transmission electron microscopy coupled with energy-dispersive X-ray spectroscopy (STEM-EDX) provides spatially resolved chemical information and is invaluable for mapping oxide-layer architecture and identifying candidate Ti/Nb-rich grain boundary phases, but in heavily alloyed superalloys it is hindered by peak overlap and limited sensitivity to dilute light elements such as boron and nitrogen \cite{Newbury2013IsQuantitative,WATANABE2006TheMethods,Keutgen2021SolvingData}. Atom probe tomography (APT) provides near-atomic-scale three-dimensional composition. However, nitrogen quantification in APT is complicated by molecular-ion formation, mass-to-charge overlaps and under-reporting, meaning that confident phase identification cannot rely on atom probe inferred chemistry alone \cite{Gault2021AtomTomography,Russo2018CompositionalEvents,Thuvander2011QuantitativeCarbides}. This is especially problematic when distinguishing nitrides with similar chemistries or when assigning crystallographic structure from compositional counting statistics. Nanoscale structural characterisation is therefore essential. 4D-STEM can provide this missing structural information by resolving local diffraction from nanoscale reaction products. A correlative approach is therefore required, combining the structural sensitivity of 4D-STEM with the three-dimensional chemical sensitivity of APT \cite{Herbig2015CombiningTomography,Ophus2019Four-DimensionalBeyond,Herbig2018SpatiallyPerspectives,Supple2023TowardsTomography}.

For interfacial problems this correlative approach is particularly powerful when combined with quantitative interface analysis. APT enables construction of proximity histograms, or proxigrams, and calculation of Gibbsian interfacial excess, providing a route to quantify elemental segregation at nitride/matrix interfaces. This is important because visual enrichment alone does not distinguish between true interfacial excess, diffuse partitioning, local depletion caused by nitride growth, or reconstruction artefacts. In systems where multiple solutes redistribute simultaneously, reconstructed atom maps, one-dimensional concentration profiles, proxigrams, Gibbsian excess analysis and local diffraction data should be interpreted together. 

Here we investigate B-containing and B-free variants of a polycrystalline Co/Ni-base superalloy to determine how boron influences oxidation-induced internal nitridation. Comparative oxidation experiments establish the effect of boron on grain boundary nitride morphology, while site-specific correlative 4D-STEM and APT are used to resolve the crystallographic structure, composition and interfacial segregation behaviour of an individual internal nitride. The reaction product is identified as a Ti/Nb-rich B1 nitride, with APT and density-corrected Gibbsian excess analysis revealing boron segregation to the nitride--matrix interface. These observations provide evidence that boron suppresses internal nitride growth by modifying the nitride-matrix interface. More broadly, the results establish interfacial solute control of internal nitridation as a strategy for designing oxidation-tolerant superalloys and chemically complex high-temperature alloys for critical aerospace applications.

\section{Results and Discussion}
\subsection{Initial Microstructure}
\begin{figure*}[!t]
\centering
 \includegraphics[width=\textwidth]{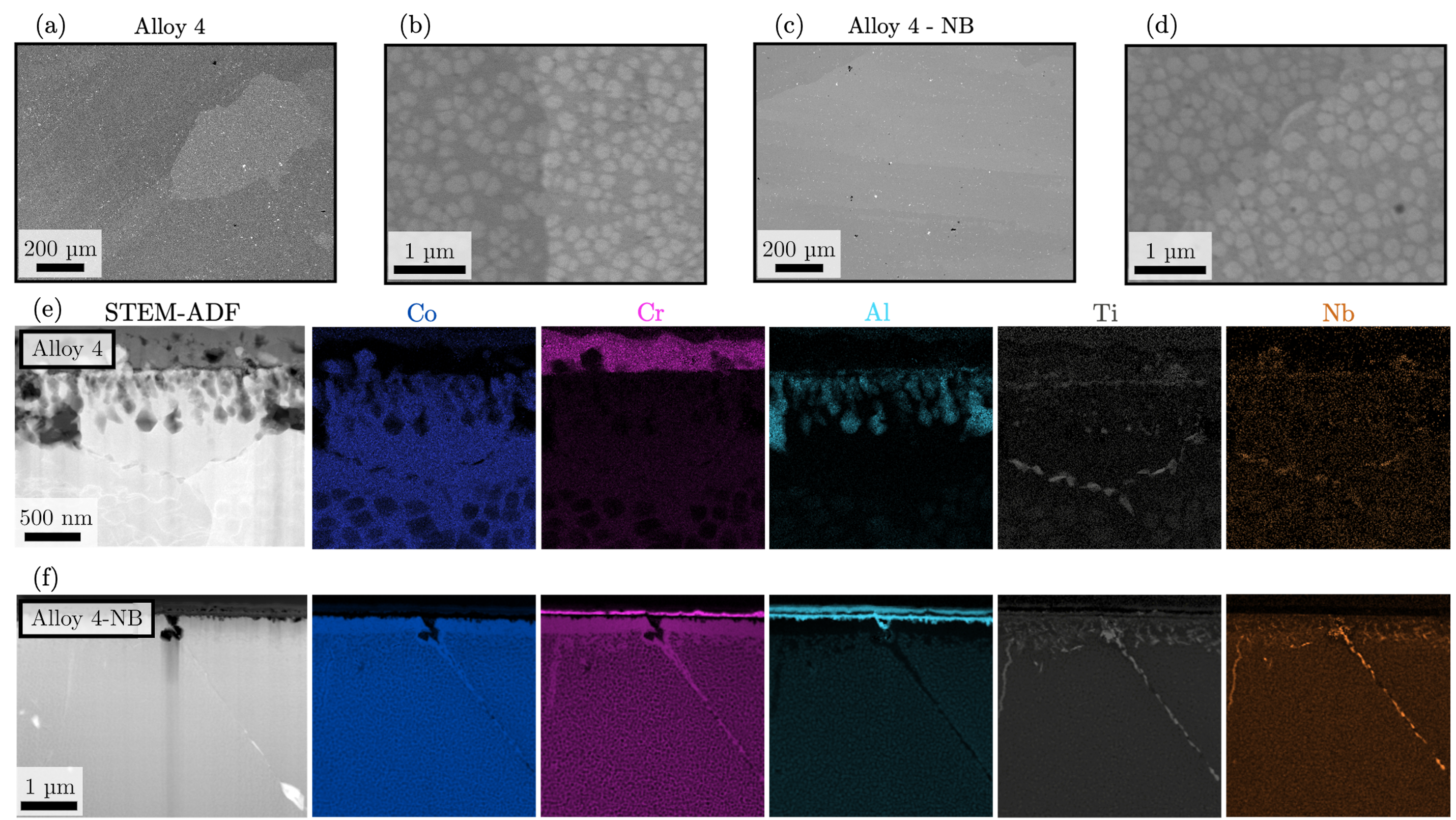}
 \caption{Initial microstructures and oxidation response of the B-containing and B-free alloys. (a, b) Initial grain structure and $\gamma'$ precipitate distribution of B-containing Alloy 4. (c, d) Corresponding microstructures of B-free Alloy 4-NB. (e, f) Cross-sectional STEM--ADF images and corresponding EDX elemental maps of oxide scales and subsurface grain boundary reaction products in Alloy 4 and Alloy 4-NB, respectively, after oxidation at \SI{800}{\degreeCelsius} for \SI{100}{\hour}. The B-free alloy shows larger Ti/Nb-rich grain boundary particles extending further beneath the oxide scale.}
  \label{fig:1}
\end{figure*}

Two compositionally matched polycrystalline Co/Ni-base superalloys were investigated: a B-containing alloy, referred to as Alloy~4, and a B-free variant, referred to as Alloy~4-NB, from \cite{Hardy2024US11898228B2}. The initial microstructure of both alloys can be seen in \textbf{Figure~\ref{fig:1}}. Both alloys were fabricated using the same processing route and differed only in boron content, as described in Section~\ref{sec:materials-processing-oxidation}. They exhibit a similar grain size of approximately \SI{500}{\micro\metre}, straight grain boundaries, and comparable $\gamma'$ distributions and fractions. This initial microstructure allows us to confidently attribute any differences in oxidation performance to the boron composition.

\subsection{Oxidation behaviour and internal nitride formation}

The B-containing and B-free variants were oxidised in static air at \SI{800}{\degreeCelsius} for \SI{100}{\hour} to compare their macroscopic oxidation behaviour and the development of internal grain boundary damage. Cross-sectional STEM--EDX was used to determine how the oxide-scale architecture and internal reaction products differ between the two alloys, as shown in \textbf{Figure~\ref{fig:1}}.

Both alloys develop a chemically stratified oxide scale in air at \SI{800}{\degreeCelsius}. The scale consists of an outer Co/Ni-rich oxide layer, consistent with spinel-type oxide formation, although it appears more pronounced on the B-containing alloy, followed by Cr- and Al-rich oxide layers beneath. This layered structure reflects the expected selective oxidation sequence in complex Ni- and Co/Ni-base superalloys widely reported in the literature, where the most mobile and reactive species oxidise according to the local oxygen potential beneath the growing scale. The Cr-rich region forms a more continuous chromia-based layer, while the Al-rich oxide appears as a thinner, less continuous sublayer with internal Al-rich protrusions extending beneath the main scale. Ti-rich oxides are also observed locally within the scale, indicating that refractory and minor reactive elements participate in the oxidation response. Previous work on this alloy family has shown that boron influences the development of the Al-rich sublayer and can redistribute to alumina-rich fingers within the oxide scale \cite{Crabb2026TimeDependentSuperalloy}.

The B-free alloy, in comparison, develops a less protective oxidation response with a more internally degraded scale morphology. The response of the B-free alloy shows significant Ti and Nb ingress. The presence of both Ti and Nb in the oxide scale in this way has been regularly documented in the literature as being deleterious to oxide performance \cite{Cruchley2013ChromiaTitanium}. However, the most significant distinction is the presence of Ti- and Nb-rich particles along the grain boundaries, often classified as nitrides or carbonitrides in the literature. Associations have been drawn between these precipitates and accelerated crack initiation \cite{Taylor2022OxidationElements,Texier2016CrackInclusions}.

The B-free alloy displays deeper and larger grain boundary nitrides, with precipitates observed at an average depth of \SI{9.7}{\micro\metre} compared to only \SI{6.1}{\micro\metre} in the B-containing alloy. Furthermore, the nitrides in the B-free alloy appear much more elongated along the grain boundary compared to the B-containing alloy, indicating that boron not only influences external oxide scale development but also plays a role in controlling the growth and morphology of internal nitrides. In this sense, boron appears to modify the internal oxidation ``damage'' mechanism itself, suppressing the development of coarse Ti/Nb-rich nitrides at grain boundaries.  Suppressing their growth therefore provides a mechanism by which boron can improve oxidation damage tolerance. To determine whether this difference is representative rather than local, the grain boundary nitride morphology was then quantified statistically.

However, Ti and N signals cannot be separated at the nanoscale using STEM--EDX: the N K$\alpha$ line (\SI{0.392}{\kilo\electronvolt}) is coincident with Ti L$\ell$ (\SI{0.395}{\kilo\electronvolt}) and overlaps Ti L$\alpha$ (\SI{0.452}{\kilo\electronvolt}), separations which are below the detector energy resolution, and nitrogen has no alternative emission line. The refractory additions suffer similar issues. Ta M$\alpha$ (\SI{1.710}{\kilo\electronvolt}), Si K$\alpha$ (\SI{1.740}{\kilo\electronvolt}) and W M$\alpha$ (\SI{1.775}{\kilo\electronvolt}) form an unresolvable triplet, while the alternative Ta L$\alpha$ (\SI{8.146}{\kilo\electronvolt}) and W L$\alpha$ (\SI{8.398}{\kilo\electronvolt}) lines straddle the intense Ni K$\beta$ line (\SI{8.265}{\kilo\electronvolt}) of the matrix. Nb L$\alpha$ (\SI{2.166}{\kilo\electronvolt}) is separated from Mo L$\alpha$ (\SI{2.293}{\kilo\electronvolt}) by only \SI{127}{\electronvolt}, comparable to the detector resolution at this energy. Projection through the TEM lamella further complicates the localisation of these grain boundary signals. This represents a common issue in reliable chemical and phase identification. STEM--EDX therefore identifies the chemical association of the internal particles with a degree of reliability, but it cannot unambiguously determine their crystallographic structure or quantify trace segregation at the interfaces, which may play a key role in the nitride growth and development. This motivates the subsequent correlative 4D-STEM and APT analysis, where local diffraction is used to determine nitride structure and atom probe tomography is used to resolve three-dimensional nitride composition and interfacial boron segregation.

\subsection{Quantification of grain boundary nitride morphology}

\begin{figure}[tp]
\centering
 \includegraphics[width=0.98\columnwidth]{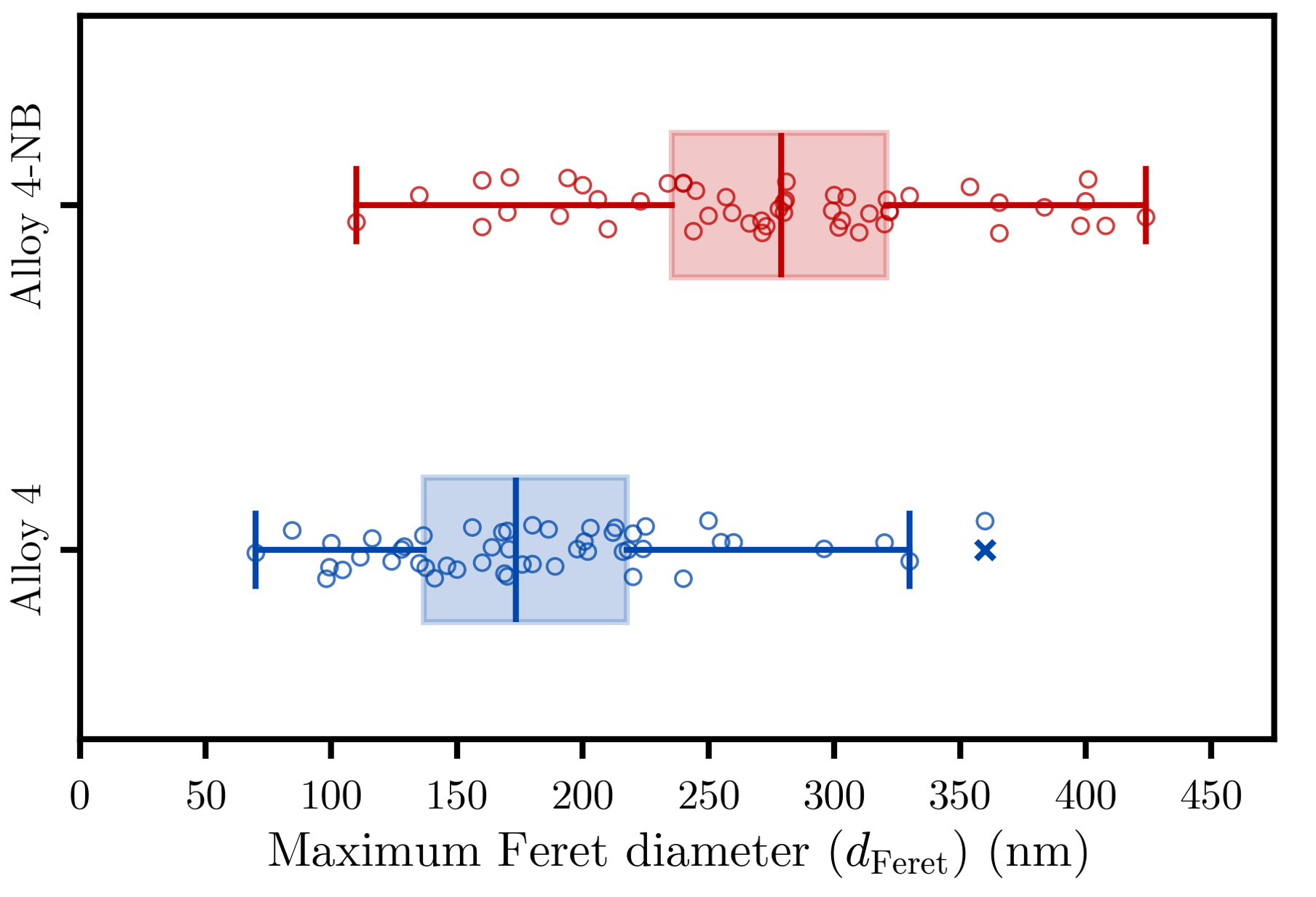}
  \caption{Distribution of internal grain boundary nitride sizes in the B-containing and B-free alloys, measured using the maximum Feret diameter, $d_{\mathrm{Feret}}$. Each distribution contains measurements from 50 nitrides across 10 grain boundaries per alloy. Boxes span the interquartile range, centre lines indicate the median, and whiskers show the full measured range.}
  \label{fig:3}
\end{figure}

To determine whether the STEM--EDX observations were representative, the grain boundary nitride population was quantified in both alloys. Nitrides were segmented from cross-sectional microscopy datasets and measured for maximum Feret diameter ($d_{\mathrm{Feret}}$) and penetration depth beneath the oxide scale, as shown in \textbf{Figure~\ref{fig:3}} and \textbf{Table~\ref{tab:nitride_summary}}.
\begin{table}[htbp]
\centering
\caption{Summary of nitride penetration-depth and size measurements. Values are
reported as the mean \(\pm\) one standard deviation (penetration depth:
\(n=10\) grain boundaries per alloy; nitride size: \(n=50\) nitrides
per alloy).}
\label{tab:nitride_summary}
\begin{tabular}{lcc}
\hline
Alloy &
{Nitride depth (\si{\micro\metre})} &
{Nitride size (\si{\nano\metre})} \\
\hline
Alloy 4    & \num{6.1 +- 0.3} & \num{180 +- 60} \\
Alloy 4-NB & \num{9.7 +- 1.6} & \num{280 +- 75} \\
\hline
\end{tabular}
\end{table}

The B-free alloy exhibits a clear shift towards larger internal nitrides. The mean nitride diameter is greater than in the B-containing alloy. The penetration depth measurements show a similar trend. In the B-free alloy, nitrides extend further beneath the oxide scale and along deeper grain boundary paths, with an average penetration depth of \SI{9.7}{\micro\metre}, indicating more extensive internal nitridation. In contrast, the B-containing alloy shows a more spatially restricted nitride population, with finer particles concentrated closer to the oxidised surface and an average penetration depth of \SI{6.1}{\micro\metre}. Together, the size and depth statistics show that boron suppresses both nitride growth and the progression of internal nitriding along grain boundaries. Therefore, boron does not prevent nitride formation altogether but instead changes the response by limiting the growth of coarse Ti/Nb-rich nitrides.

\subsection{Correlative 4D-STEM and APT analysis of internal nitrides}

To determine the crystallographic structure and three-dimensional chemistry of the internal nitrides, a representative grain boundary particle from the B-containing alloy was selected for correlative 4D-STEM and APT analysis, following the workflow described in the Methods section. The region was located beneath the external oxide scale within an oxidised grain boundary region, consistent with the Ti/Nb-rich internal nitrides identified by STEM-EDX. This analysis was required because the literature on internal nitridation in superalloys has most commonly assigned such reaction products from chemistry alone, typically as Ti- or Al-rich nitrides/carbonitrides, while direct nanoscale structural identification remains uncommon \cite{Taylor2022OxidationElements,Krupp1999InternalSystem,Krupp1999InternalDescription}. 

\begin{figure*}[!t]
\centering
 \includegraphics[width=0.82\textwidth]{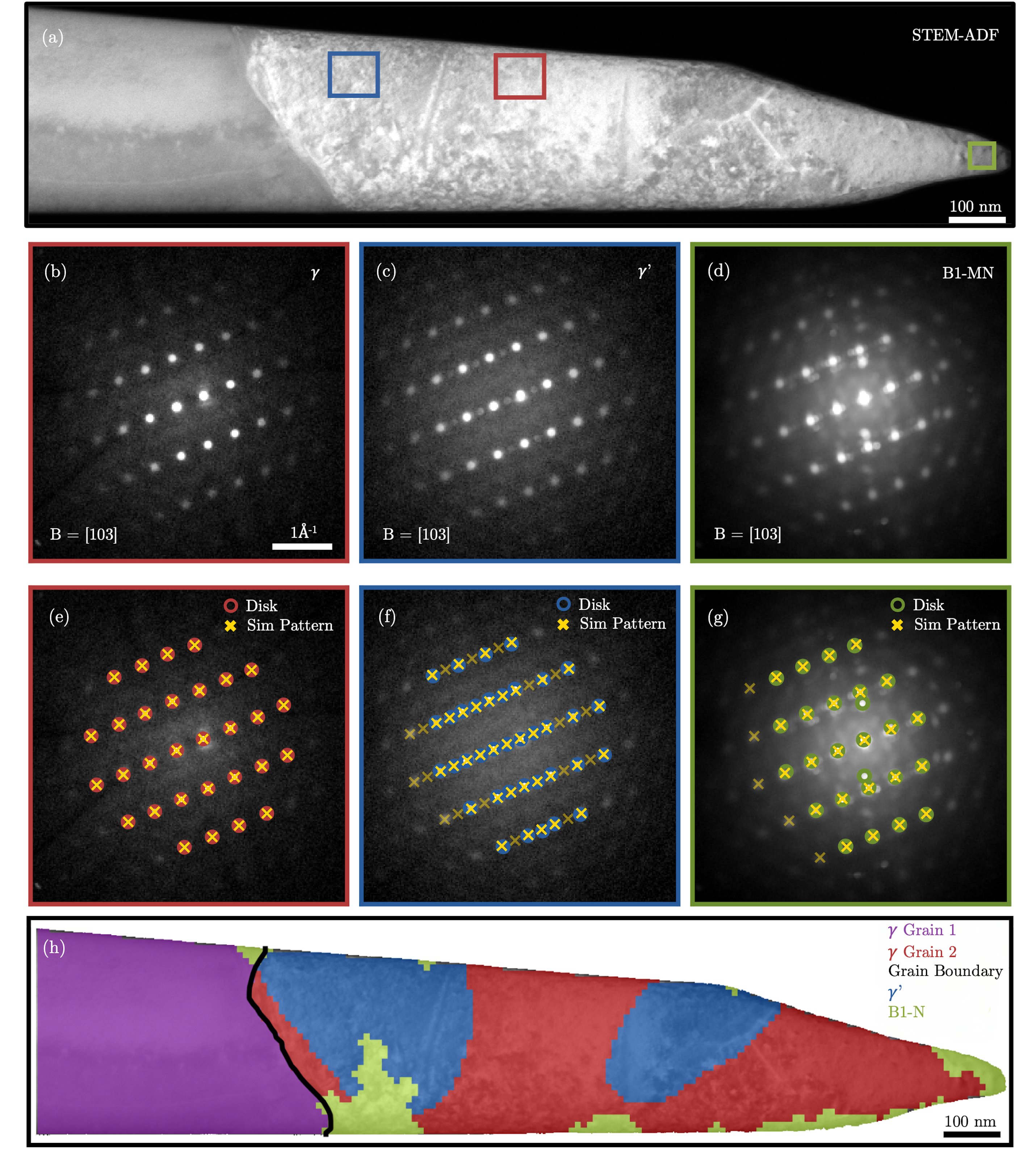}
  \caption{Correlative 4D-STEM analysis of the APT needle containing an internal grain boundary nitride. 
(a) Virtual ADF image showing the analysed needle and regions selected for diffraction analysis. 
(b--d) Representative diffraction patterns obtained from the highlighted regions. 
(e--g) Detected diffraction disks overlaid with simulated patterns used for phase identification. 
(h) Integrated diffraction-based phase map separating the $\gamma/\gamma^\prime$ matrix from the B1 rock-salt-type nitride.}
  \label{fig:4DSTEM_Figure}
\end{figure*}

\textbf{Figure~\ref{fig:4DSTEM_Figure}} shows the STEM-ADF image of the analysed region, representative diffraction patterns from selected regions of the needle, automated disk detection used to match against potential phases, and the resulting diffraction-based phase map. The phase map separates the surrounding $\gamma/\gamma'$ matrix from the internal grain boundary reaction product using local diffraction information, rather than projected chemical contrast. Diffraction patterns from the $\gamma$ and $\gamma'$ regions provide internal references for the local orientation and spacing calibration, while the particle produces a distinct diffraction signature consistent with a B1 rock-salt-type nitride structure. This is important because B1 nitrides such as TiN and NbN are structurally similar and can be difficult to distinguish from diffraction geometry alone, particularly in nanoscale, strained and dynamically scattering specimens \cite{Toth1971,Ophus2019Four-DimensionalBeyond,Pekin2017OptimizingMapping}. The conclusion from 4D-STEM is therefore that the grain boundary reaction product is a crystallographically distinct B1-type nitride. Furthermore, we can confirm this by calculating the lattice parameter, a, through measuring the reciprocal lattice spacing after calibration. The resulting lattice parameter for the B1 phase identified here is 
\(a_{\mathrm{B1}}=4.23\pm0.08~\si{\angstrom}\). This is close to the nominal lattice parameter of TiN, 
$a_{\mathrm{TiN}} \approx \SI{4.24}{\angstrom}$, and lower than that of NbN, $a_{\mathrm{NbN}} \approx \SI{4.39}{\angstrom}$, supporting assignment as a Ti-rich B1-type nitride \cite{Lengauer2000HandbookMaterials}. The matrix and nitride atomic number densities were calculated from the lattice parameters measured by 4D-STEM rather than from the reconstructed APT density, which is affected by local magnification and trajectory aberrations. Because N is under-counted in APT, the local nitride fraction used in the density correction was determined from Ti+Nb relative to all consistently ranged alloy cations. N was used only to construct the continuous reference profile in Section~2.5 due to the strong partitioning to the nitride phase. This calculation gave an equivalent precipitate diameter of \SI{21.5 \pm 0.4}{\nano\metre}, consistent with its approximate \SI{20}{\nano\metre} reconstructed size. The benefit of the correlative approach is that 4D-STEM identifies the B1 structure of the nitride while APT establishes the Ti-dominated, Nb-containing composition. These measurements provide the structural and compositional inputs required for the density-corrected Gibbsian excess calculations in Section~2.5.

This structural assignment advances the interpretation beyond previous studies where internal nitrides beneath oxide scales were commonly reported as TiN, NbN or AlN primarily from SEM or STEM-EDX alone. In the present alloy, the structural identification of a B1-type nitride is significant; however, 4D-STEM cannot fully resolve the cation chemistry of TiN and NbN family nitrides because both adopt related rock-salt structures and have similar diffraction spacings. The structural result must therefore be combined with composition obtained from APT to determine whether the nitride is Ti-rich, Nb-rich or chemically mixed.

\begin{figure*}[!t]
\centering
 \includegraphics[width=\textwidth]{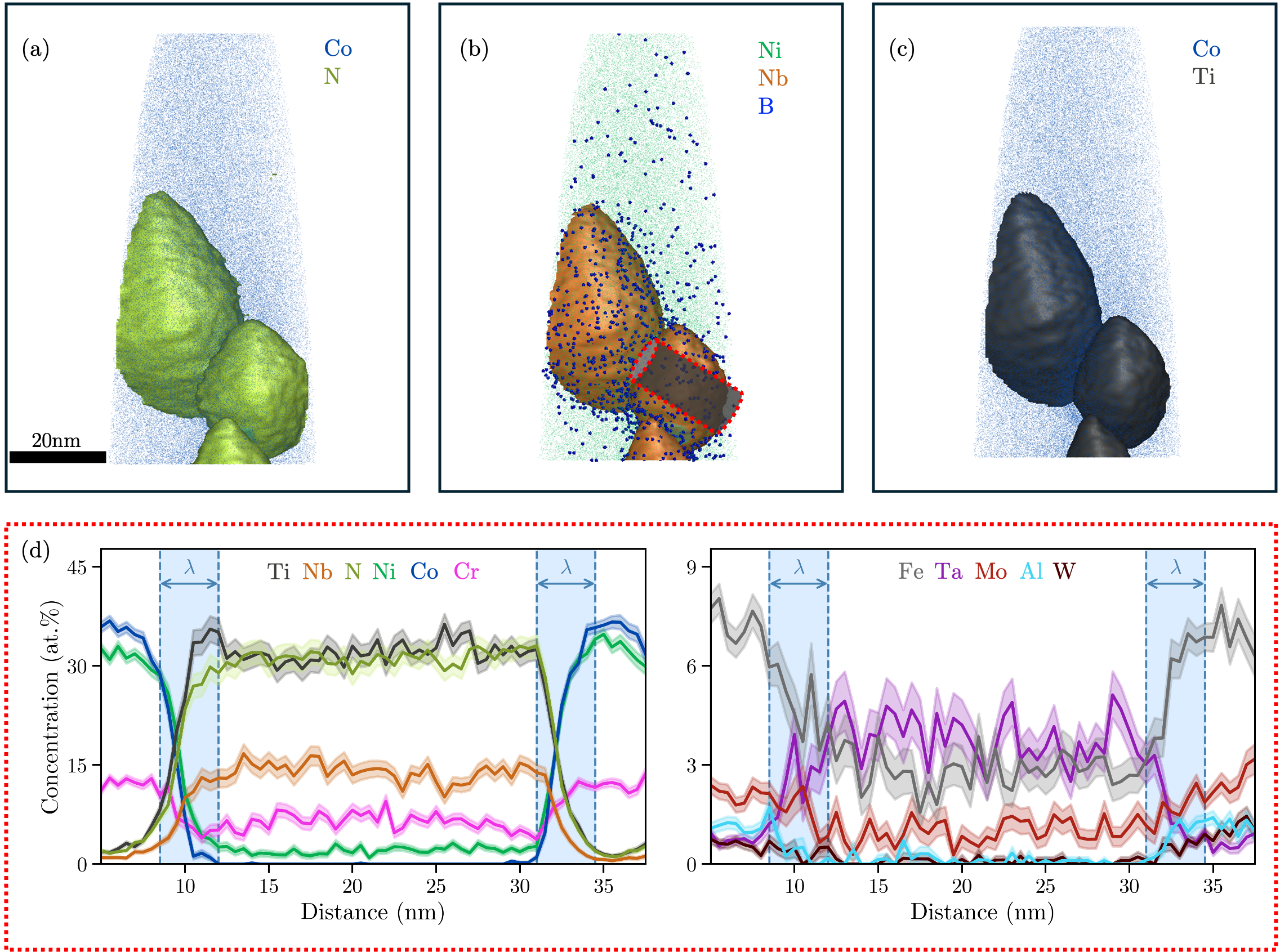}
  \caption{Atom probe tomography analysis of the correlative needle containing the internal nitride. 
(a--c) Reconstructions showing the spatial distribution of Ti, Nb and N within the nitride, with (a) Co, N, (b) Ni, Nb, B, (c) Co, Ti. 
(d) One-dimensional concentration profiles across the nitride--matrix interface, where $\lambda$ denotes the interface width. Shaded bands around the concentration profiles represent $\pm 1$ standard error from APT counting statistics.}
  \label{fig:APT_Figure}
\end{figure*}

Correlative APT analysis of the same site in \textbf{Figure~\ref{fig:APT_Figure}} confirms that the B1-structured particle is a Ti-dominated nitride, with significant Nb enrichment in addition to other contributions from Cr and Ta. One-dimensional concentration profiles through the particle show strong Ti, Nb and N enrichment relative to the surrounding $\gamma/\gamma'$ matrix, with corresponding depletion of Ni, Co and Al. This indicates that the analysis is consistent with the formation of a chemically complex internal nitride within a multicomponent superalloy rather than a pure binary nitride as previously thought. This agrees with the broader understanding that internal precipitates can incorporate minor solutes or develop local compositional gradients during growth \cite{Krupp1999InternalDescription,Cruchley2013ChromiaTitanium}. 

The absence of significant C or O enrichment supports assignment of the particle as a nitride rather than a carbonitride or oxide. However, the measured nitrogen concentration is lower than the 50\,at.\% expected for a B1-type mononitride. Nitrogen was ranged as both elemental and molecular ion species, with the principal mass-to-charge overlaps resolved by isotopic decomposition. With these overlaps accounted for, the nitrogen concentration remains below the stoichiometric value. Accurate nitrogen quantification in APT is a widely reported issue, particularly under laser-pulsed conditions, with deviations linked to molecular-ion formation, dissociation behaviour, multiple-hit detection losses and possible neutral losses \cite{Russo2018CompositionalEvents, Gault2016BehaviorEmitter}. The residual under-counting is therefore attributed to these detection limitations rather than evidence against nitride formation. Because the B1 structure determined by 4D-STEM fixes the anion sublattice at 50\,at.\%, the cation composition can be assessed independently of the nitrogen under-counting by normalising over the metal sublattice only, confirming a Ti-dominated, Nb-enriched mixed nitride.

Significantly, APT shows clear boron enrichment localised at the nitride/matrix interface rather than throughout the nitride volume or distributed randomly in the matrix. The interfacial enrichment is present independently in both the B$^{+}$ and B$^{2+}$ charge states, confirming it is not an artefact of field-dependent redistribution between ion species across the interface. This observation is consistent with the known tendency of boron to be a segregator in superalloys, but extends that behaviour from conventional grain boundary segregation to oxidation-induced nitride interfaces. This result suggests that boron interacts directly with the nitride boundary, providing a possible explanation for the suppressed nitride growth observed statistically in the B-containing alloy. However, because apparent interfacial enrichment in APT can be affected by diffuse interfaces, density changes, reconstruction artefacts and lateral non-uniformity of segregation across the interface \cite{Jenkins2020ReflectionsTomography}, this boron enrichment must be quantified before it can be interpreted mechanistically. The following section therefore applies a density-corrected Gibbsian excess analysis to determine whether the observed boron peak represents genuine interfacial segregation. 

Together, these results highlight why a correlative 4D-STEM/APT approach is required to resolve complex mechanisms in complex alloy systems. 4D-STEM establishes that the grain boundary reaction product is structurally a B1-type nitride, while APT confirms that the composition is $(\mathrm{Ti},\mathrm{Nb},X)\mathrm{N}$, in addition to revealing local boron enrichment at the nitride/matrix interface. This combined structural and compositional evidence links the statistical suppression of nitride growth to a specific interfacial mechanism: boron segregation at the internal nitride--matrix boundary.
\subsection{Gibbsian excess analysis of the nitride--matrix interface}

The APT reconstruction shows clear boron enrichment at the nitride--matrix interface; however, visual enrichment in an atom probe reconstruction is not, by itself, sufficient evidence for true interfacial segregation. This distinction is central to quantitative APT analysis. Proximity histograms report the average composition within shells at fixed distances from an isoconcentration surface, summed over the entire surface, meaning that locally enriched or depleted regions are averaged into a single value at each distance. The resulting profiles are additionally sensitive to interface definition, local magnification, trajectory aberrations and reconstruction broadening \cite{Gault2021AtomTomography,Vurpillot2000TrajectoryProbe,Hellman2000AnalysisHistogram,Vurpillot2004ImprovementProbe,Grenier20143DTomography}. Where multiple interfaces lie in close proximity, shells extended from one surface can also sample solute redistribution associated with neighbouring features. The proxigram analysis range was therefore restricted to distances over which the shells did not intersect the adjacent nitride particle or its associated solute-redistribution zone, and the averaged profiles were cross-checked against the independent one-dimensional profiles in \textbf{Figure 4d}.

This is particularly important for this example because the nitride and surrounding matrix have different compositions, different atomic number densities, and a non-zero interfacial width. In this situation, a conventional proxigram combines several effects: genuine segregation, diffuse partitioning between phases, local depletion caused by nitride growth, and reconstruction-related broadening. A quantitative Gibbsian excess analysis was therefore applied to determine whether the boron peak represents true interfacial excess rather than an artefact of the interface definition.

The nitride--matrix interface was defined using a 3~at.\% N isoconcentration surface calculated from the total decomposed nitrogen count, which captures the boundary of the grain boundary nitride directly. Proximity histograms were then generated normal to this surface. This interface definition is appropriate because nitrogen is the defining species of the precipitate, whereas Ti and Nb may vary locally depending on the cation distribution within the particle, as well as being present in the alloy prior to internal nitriding. Defining the interface using nitrogen therefore avoids biasing the analysis towards either Ti-rich or Nb-rich regions and provides a chemically meaningful boundary for quantifying segregation at the nitride--matrix interface. Because interfacial excess values can depend on the selected isoconcentration threshold, as shown for proxigram-based APT analysis by Yoon et al. \cite{Yoon2004DependenceSurface}, the robustness of the interface position was assessed by varying the nitrogen isovalue within a reasonable range and confirming that the boron peak remained localised at the interface.

\begin{figure*}[!t]
\centering
 \includegraphics[width=\textwidth]{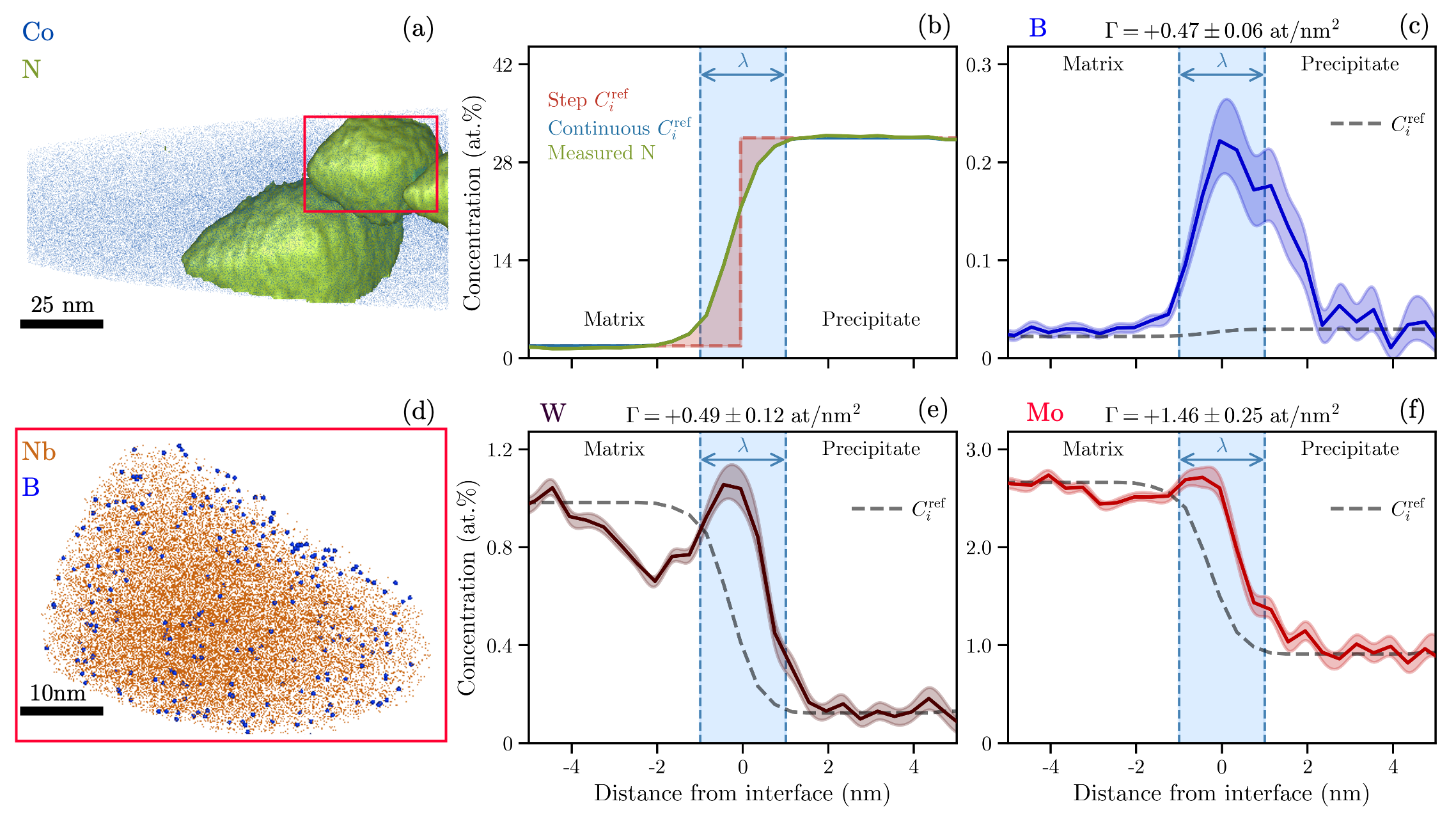}
 \caption{Density-corrected Gibbsian interfacial excess analysis of the nitride--matrix interface. A \SI{3}{at.\percent} N isoconcentration surface was used to define the nitride interface.
(a) APT reconstruction showing the nitride isosurface. 
(b) Comparison between the conventional step-function reference and the continuous nitrogen-based reference profile. 
(c,e,f) Concentration profiles and corresponding Gibbsian interfacial excesses for B, W and Mo, respectively, where $\lambda$ denotes the interface width. 
(d) Reconstruction showing Nb enrichment within the nitride and B segregation at the nitride--matrix interface. Shaded bands around the concentration profiles represent $\pm 1$ standard error from APT counting statistics. Quoted uncertainties in the Gibbsian interfacial excesses represent one standard deviation incorporating both APT counting statistics and uncertainties in the matrix and nitride lattice parameters.}
  \label{fig:Gibbsian_Excess}
\end{figure*}

A conventional Gibbsian excess analysis treats the interface as a sharp Gibbs dividing surface and calculates excess relative to a step-function reference profile. While this approach is appropriate for chemically abrupt interfaces, it is less suitable for the present nitride--matrix boundary because the interface is compositionally diffuse within the reconstruction and the adjoining phases have different atomic number densities. As shown in Figure~\ref{fig:Gibbsian_Excess}b, a step-function reference can therefore mis-assign gradual matrix--nitride partitioning as interfacial excess. To avoid this, a density-corrected Gibbsian excess approach was used, with the full derivation and implementation described in the Methods section. Briefly, the decomposed nitrogen profile was used to construct a continuous reference profile for each element, while the local atomic number density was allowed to vary across the interface using matrix and B1 nitride densities calculated from the 4D-STEM-measured lattice parameters. The final density-corrected excess was then calculated as

\begin{equation}
\Gamma_i =
\int_{z_{\mathrm{min}}}^{z_{\mathrm{max}}}
n(z)
\left[
C_i(z)-C_i^{\mathrm{ref}}(z)
\right]
\,\mathrm{d}z,
\label{eq:gibbs_density_corrected}
\end{equation}

where \(C_i(z)\) is the measured proxigram concentration profile, \(C_i^{\mathrm{ref}}(z)\) is the continuous nitrogen-based reference profile, and \(n(z)\) is the local atomic number density across the nitride--matrix interface. The resulting \(\Gamma_i\) gives the density-corrected interfacial excess of species \(i\), with units of \(\mathrm{atoms\,nm^{-2}}\) when \(n(z)\) is given in \(\mathrm{atoms\,nm^{-3}}\) and \(z\) is given in \(\mathrm{nm}\).

\textbf{Figure~\ref{fig:Gibbsian_Excess}} summarises this analysis. The APT reconstruction defines the Ti/Nb-rich nitride and the nitrogen isoconcentration surface used to construct the proxigram. The measured profiles show strong Ti, Nb and N enrichment inside the nitride, depletion of matrix elements across the interface, and a narrow boron peak at the nitride--matrix boundary. The comparison between the step-function reference and the continuous nitrogen-based reference highlights why the reference choice matters: a sharp step function can overestimate or mis-assign interfacial excess for elements whose concentration changes gradually across the nitride boundary, whereas the continuous reference separates diffuse partitioning from genuine interfacial enrichment.

The density-corrected Gibbsian excess analysis confirms positive interfacial excess for B, W and Mo at the nitride--matrix boundary, as summarised in \textbf{Figure~\ref{fig:Gibbsian_Excess}}. The positive boron excess demonstrates that the boron peak observed in the proxigram is not simply caused by interface broadening, phase partitioning, or the choice of isoconcentration surface. Instead, boron is genuinely segregated to the interface between the Ti/Nb-rich nitride and the surrounding matrix. W and Mo also show positive residual enrichment relative to the continuous nitrogen-based reference; however, this enrichment is interpreted differently from boron. Because W and Mo are refractory matrix solutes with limited incorporation into the B1 nitride, their peak-and-dip profiles are consistent with solute rejection and pile-up at a moving nitride growth front, rather than simple equilibrium segregation to a fixed interface. The nitride interface is therefore associated with local redistribution of refractory solute chemistry as well as boron segregation. Whether these features are present around the whole interface, rather than only in the averaged proxigram, was checked using independent one-dimensional concentration profiles at multiple locations around the nitride--matrix interface. The B enrichment and the W and Mo peak-and-dip shape appear in each individual profile, confirming they are consistent features of the interface itself rather than composites of dissimilar regions, such as genuine enrichment in one region averaged with preferential-retention tails in another \cite{Jenkins2020ReflectionsTomography}.
The positive boron excess provides the mechanistic link between the APT chemistry and the nitride morphology statistics. Boron is known to segregate to grain boundaries and other high-energy interfaces in superalloys, where it can reduce interfacial energy and improve cohesion \cite{Kontis2016OnSuperalloy}. The present analysis shows that this behaviour extends to oxidation-induced nitride interfaces. By segregating to the nitride--matrix boundary, boron may reduce the interfacial energy of the growing nitride, suppressing coarsening and limiting the development of larger, more deleterious grain boundary nitrides. This interpretation is consistent with the statistical observation that the B-containing alloy contains finer and less extensive internal nitrides than the B-free alloy.

The Gibbsian excess analysis therefore provides the quantitative basis for the proposed growth-suppression mechanism. The morphology statistics show that boron suppresses coarse nitride formation and reduces penetration; correlative 4D-STEM/APT identifies the particles as Ti/Nb-rich B1 nitrides; and the density-corrected excess analysis demonstrates that boron segregates specifically to the nitride--matrix interface. Together, these results show that boron modifies the interfacial chemistry of oxidation-induced nitrides, providing a nanoscale mechanism for the reduced internal nitride damage observed in the B-containing alloy.

\subsection{Mechanism of boron-mediated nitride growth suppression}

The results support a mechanism in which boron suppresses grain boundary nitrogen degradation by modifying the interface of growing internal nitrides. The key observation is not simply that the B-containing alloy forms smaller nitrides, but that boron is quantitatively enriched at the nitride--matrix interface where nitride growth and advance must occur. This links the statistical suppression of coarse and deeply penetrating grain boundary nitrides to a specific nanoscale chemical feature: a boron-decorated B1 nitride interface.

\begin{figure*}[!t]
\centering
 \includegraphics[width=\textwidth]{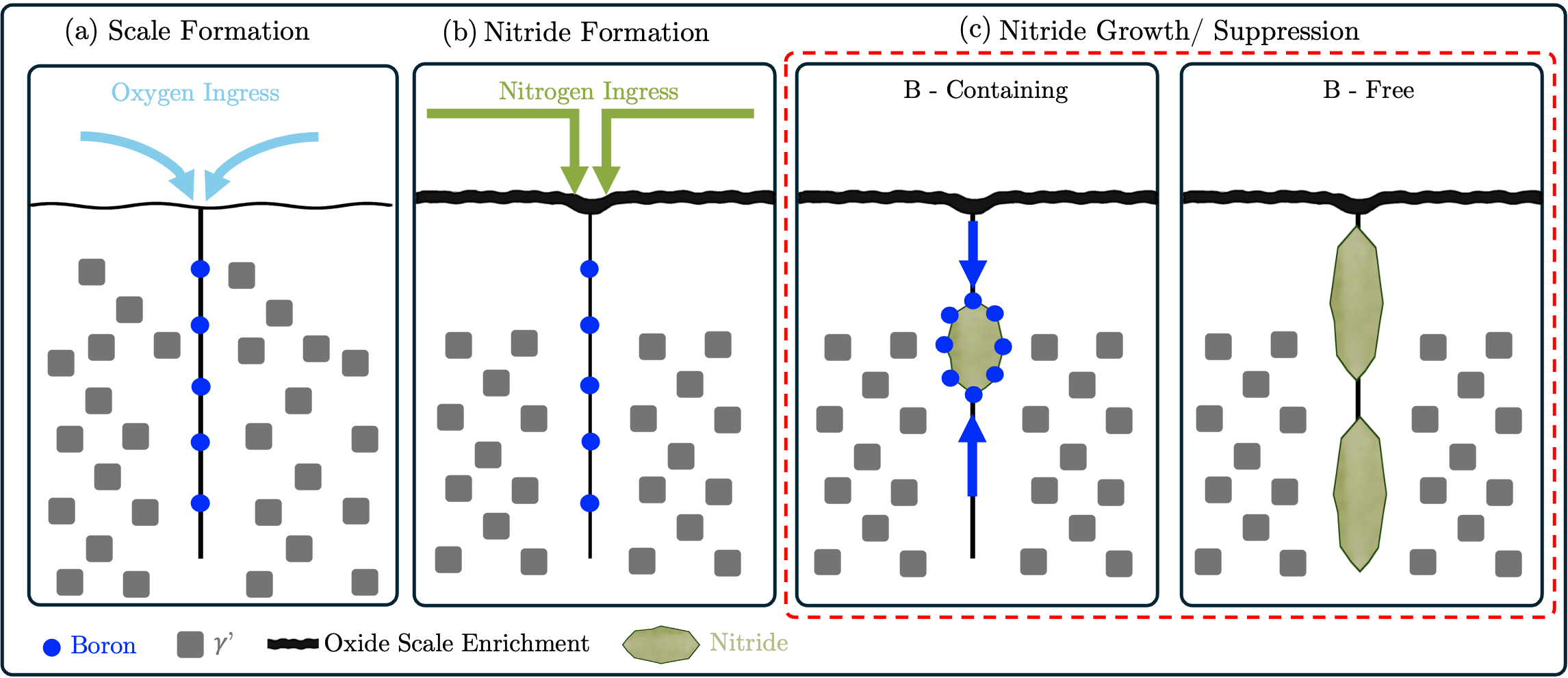}
  \caption{Schematic illustration of the proposed mechanism for boron-mediated suppression of internal nitride growth. 
(a) Initial oxide-scale formation during exposure in air. 
(b) Nitrogen ingress beneath the oxide scale as the local oxygen potential decreases. 
(c) Comparison of nitride growth and penetration in the B-containing and B-free alloys.}
  \label{fig:Mechanism}
\end{figure*}

The proposed sequence is summarised in \textbf{Figure~\ref{fig:Mechanism}}. During exposure at \SI{800}{\degreeCelsius} in air, selective oxidation produces a chemically stratified external scale, while oxygen consumption beneath the scale lowers the local oxygen potential and enables nitrogen ingress into the alloy. This is consistent with established descriptions of internal nitridation in Ni-base superalloys, where nitrogen can penetrate beneath damaged or imperfectly protective scales and where Ti-, Nb- and Al-rich nitrides are commonly reported beneath chromia- and alumina-containing oxide layers.

In the present alloy, this internal-nitridation pathway is modified by boron. STEM--EDX and statistical analysis show that both the B-containing and B-free alloys form internal grain boundary nitrides after oxidation, demonstrating that boron does not completely prevent nitrogen ingress or suppress nitride nucleation. Instead, the dominant effect is on nitride growth and penetration. The B-free alloy develops more elongated nitrides that extend further along grain boundaries, whereas the B-containing alloy retains a finer and more spatially restricted nitride population. Boron therefore appears to limit the advance and coarsening of nitrides once internal nitridation has begun.

In this picture, boron does not prevent the initial formation of nitrides; instead, it limits their subsequent development into larger, more deeply penetrating grain boundary particles. The W and Mo profiles support this interpretation by showing local refractory-solute redistribution around the nitride interface. Their positive residual enrichment and adjacent depletion are consistent with solute rejection and pile-up at an advancing B1 nitride growth front. The nitride--matrix interface is therefore not a passive boundary between two phases, but an active reaction front involving nitrogen uptake, Ti/Nb enrichment, refractory-solute redistribution and boron segregation.

This mechanism has direct significance for oxidation damage tolerance. Internal nitridation is not merely a chemical by-product of exposure; it creates hard, brittle nitride particles at grain boundaries. By extending further beneath the oxide scale, these particles extend the depth of oxidation damage and the near-surface region that may show reduced strain to fracture. This could produce micro-fissures that subsequently develop into fatigue cracks, consistent with previous crack-tip and fatigue studies \cite{Kitaguchi2013OxidationSuperalloy,Cruchley2015TheSuperalloy,Texier2016CrackInclusions}. Suppression of coarse Ti/Nb-rich grain boundary nitrides is therefore directly relevant to oxidation-assisted cracking, ductility loss and component lifetime.

The broader implication is that boron should not be viewed only as a conventional grain boundary strengthener in polycrystalline superalloys. Here, boron acts as an interfacial dopant that modifies the growth of oxidation-induced internal nitrides. This reframes minor-element alloying as a means of controlling internal reaction-front chemistry beneath oxide scales, rather than only improving bulk oxidation kinetics or grain boundary cohesion. Although demonstrated here in a Co/Ni-base superalloy, the same principle is relevant to other chemically complex high-temperature alloys, including advanced superalloys and compositionally complex alloys, where localised oxidation damage, internal nitridation and interface stability are critical to long-term performance.

The mechanism proposed here therefore extends the conventional role of boron beyond grain boundary cohesion during creep. During oxidation, boron redistributes to the interface of newly formed internal nitrides, where it suppresses nitride growth and reduces the penetration of subsurface grain boundary damage. This provides a mechanistic basis for why minor boron additions can improve oxidation damage tolerance in polycrystalline superalloys: they reduce the development of coarse Ti/Nb-rich nitrides along grain boundaries, a damage mode linked in the literature to cracking, ductility loss and reduced lifetime.

\section{Methods}

\subsection{Materials, processing and oxidation exposure}
\label{sec:materials-processing-oxidation}
\begin{table*}[htbp]
\centering
\caption{Nominal alloy compositions in wt.\%, from \cite{Hardy2024US11898228B2}.}
\label{tab:alloy_compositions}
\renewcommand{\arraystretch}{1.0}
\setlength{\tabcolsep}{4pt}
\small
\resizebox{\textwidth}{!}{%
\begin{tabular}{lccccccccccccccc}
\toprule
        & \textbf{Ni} & \textbf{Co} & \textbf{Cr} & \textbf{Fe} & \textbf{W} & \textbf{Mo} &
          \textbf{Al} & \textbf{Ta} & \textbf{Nb} & \textbf{Ti} &
          \textbf{Mn} & \textbf{Si} & \textbf{C} & \textbf{B} & \textbf{Zr} \\
\midrule
Alloy 4    & Bal. & 28.00 & 10.50 & 5.00 & 5.39 & 2.90 & 4.15 & 3.00 & 1.50 & 1.50 & 0.50 & 0.10 & 0.015 & 0.009 & 0.038 \\
Alloy 4-NB & Bal. & 28.00 & 10.50 & 5.00 & 5.39 & 2.90 & 4.15 & 3.00 & 1.50 & 1.50 & 0.50 & 0.10 & 0.015 & --    & 0.038 \\
\bottomrule
\end{tabular}
}
\end{table*}
The nominal compositions of both alloys are given in \textbf{Table~\ref{tab:alloy_compositions}}. Both alloys were produced by vacuum induction melting (VIM) at Imperial College London using the same melt route, with ingot masses of approximately \SI{450}{\gram}. The ingots were homogenised under vacuum using a two-step heat treatment consisting of \SI{1160}{\degreeCelsius} for \SI{6}{\hour}, followed by \SI{1210}{\degreeCelsius} for \SI{66}{\hour}, before cooling in air.

Oxidation specimens for microscopy were sectioned from the homogenised ingots with dimensions of approximately \SI{10}{\milli\metre} $\times$ \SI{10}{\milli\metre} $\times$ \SI{2}{\milli\metre}. Prior to oxidation, the samples were ground and polished using standard metallographic procedures, followed by a final \SI{10}{\minute} neutralised oxide polishing suspension (OPS) polish to obtain a reproducible surface finish. One specimen from each alloy was oxidised in a separate exposure using the same furnace, in an alumina crucible under static laboratory air at \SI{800}{\degreeCelsius} for \SI{100}{\hour}. The temperature of each exposure was verified using a thermocouple.

\subsection{Nitride morphology statistics and reproducibility}

Cross-sectional scanning electron microscopy (SEM) was used to quantify the morphology and penetration depth of internal grain boundary nitrides in Alloy~4 and Alloy~4-NB after oxidation. Imaging was performed using a Thermo Fisher Scientific Helios G5~CX operated at \SI{5}{\kilo\volt} and \SI{0.69}{\nano\ampere}, using the Everhart--Thornley secondary electron detector (ETD). Images were acquired at a stage tilt of \SI{52}{\degree}, and image dimensions were corrected for the tilt geometry prior to quantitative analysis.

For each alloy, nitride size was measured for \(n=50\) nitrides sampled across evenly distributed locations across \(n=10\) grain boundaries in the oxidised specimens. Penetration
depth was measured once for each of the same grain boundaries (\(n=10\) per alloy). No nitrides or grain boundaries were excluded. Nitride size and penetration were measured using ImageJ software. The maximum Feret diameter was used as the primary particle-size metric because it captures the elongated morphology of grain boundary nitrides more appropriately than an equivalent circular diameter.

Penetration depth was measured by following the associated grain boundary trace from the oxide/matrix interface to the deepest observed nitride along that boundary. This definition was used rather than a simple vertical distance from the surface so that inclined grain boundaries were treated consistently.

Sample sizes were selected to provide spatial coverage, without a formal statistical power calculation. The measurements were analysed descriptively and therefore represent specimen subsampling. No inferential statistical hypothesis tests were performed.

\subsection{STEM imaging and STEM--EDX}

Cross-sectional lamellae for STEM imaging were prepared from oxidised regions of interest using a dual-beam focused ion beam/scanning electron microscope (Thermo Fisher Scientific FEI Helios G5~CX). Site-specific lift-out was performed from grain boundary locations. Prior to milling, protective C and Pt layers were deposited by electron beam. The lamellae were then thinned to electron transparency using a multi-step ion milling procedure, with final thinning performed at \SI{5}{\kilo\volt} and \SI{68}{\pico\ampere} to reduce Ga implantation and FIB-induced damage.

STEM imaging was carried out using a Thermo Fisher Scientific probe-corrected, monochromated XFEG Spectra~300 operated at \SI{300}{\kilo\volt}, equipped with a double-tilt holder. A convergence semi-angle of \SI{30}{\milli\radian} was used, and annular dark-field imaging was acquired using a HAADF detector with a collection-angle range of \SIrange{52}{200}{\milli\radian}. Energy-dispersive X-ray spectroscopy was acquired using a Dual-X detector system with a probe current of \SI{1}{\nano\ampere}. STEM and EDX datasets were processed using Thermo Fisher Scientific \emph{Velox} software with drift-corrected frame integration to minimise scan distortion. The elemental maps in \textbf{Figure 1e,f} are displayed as wt.\% maps generated from the Co K$\alpha$, Cr K$\alpha$, Al K$\alpha$, Ti K$\alpha$ and Nb L$\alpha$ lines, with background correction and peak deconvolution applied in Velox. These maps are used to show the spatial distribution of each element rather than for quantitative compositional analysis.

\subsection{4D-STEM analysis}

The 4D-STEM datasets were acquired using the same Thermo Fisher Scientific Spectra~300 operated at \SI{300}{\kilo\volt}. Diffraction patterns were recorded using a Gatan K3 direct electron detector mounted after a Gatan Continuum energy filter and operated in counting mode. Data were collected in EFTEM mode at a camera length of \SI{180}{\milli\metre}. The diffraction datasets were analysed using the open-source py4DSTEM package \cite{Savitzky2021Py4DSTEM:Analysis}.

Diffraction disks were identified by cross-correlation-based peak finding, and subsequent virtual imaging, pattern comparison, phase mapping and strain-related analysis were performed from the detected disk positions. Reciprocal-space calibration for lattice-spacing measurements was performed using the Pt diffraction spacing from the protective layer. The calibrated diffraction data were used to identify the crystallographic structure of the grain boundary reaction product and to estimate the lattice parameter of the B1 nitride phase used in the density-corrected interface analysis.
\subsection{Atom probe tomography and interface analysis}

{\emergencystretch=2em \tolerance=1700 APT analysis was performed using a CAMECA LEAP 5000 XR instrument in UV laser-pulsing mode (\(\lambda = \SI{355}{\nano\metre}\)). Operating conditions comprised a laser repetition rate of \SI{200}{\kilo\hertz}, a pulse energy of \SI{50}{\pico\joule}, a specimen temperature of \SI{50}{\kelvin}, and a detection rate of \(0.01\) ions pulse\(^{-1}\).\par}

APT reconstruction and data analysis were carried out using the IVAS software within CAMECA \emph{AP Suite}~6.3. The three-dimensional reconstructions were generated using an initial tip radius and shank angle defined to match the TEM image of the same needle. Elemental distributions were examined using reconstructed atom maps, one-dimensional concentration profiles and proximity histograms. The proximity histogram used for the interfacial analysis was generated from a 3~at.\% N isoconcentration surface constructed from the total decomposed N profile. N-containing molecular ions (TiN$^{2+}$, TiN$^{+}$, NbN$^{2+}$) were decomposed into their elemental contributions before construction of the total N profile. Peak overlaps were deconvolved and background correction applied within IVAS. B was ranged as both the B$^{+}$ and B$^{2+}$ charge states consistently across the matrix, interface and nitride regions, and the nitride-interior mass spectrum contained no unranged B-containing molecular species.

In the classical Gibbsian framework, interfacial excess is calculated relative to a sharp Gibbs dividing surface using a step-function reference profile. This approach follows the application of Gibbsian interfacial excess to atom-probe measurements by Krakauer and Seidman, and its later extension to arbitrary interface geometries in three-dimensional atom-probe microscopy by Hellman and Seidman \cite{Krakauer1993AbsoluteInterfaces,Hellman2002MeasurementMicroscopy}. For a solute \(i\), the apparent excess may be written as
\begin{equation}
\Gamma_i^{\mathrm{step}} =
n_0
\int_{z_{\mathrm{min}}}^{z_{\mathrm{max}}}
\left[
C_i(z)-C_i^{\mathrm{step}}(z)
\right]
\,\mathrm{d}z,
\label{eq:gibbs_step}
\end{equation}
where \(\Gamma_i^{\mathrm{step}}\) is the apparent interfacial excess, \(n_0\) is the atomic number density, \(C_i(z)\) is the measured concentration profile, and \(C_i^{\mathrm{step}}(z)\) is the idealised step-function reference constructed from the far-field matrix and nitride compositions. This approach is appropriate for chemically sharp interfaces, but it is less suitable for diffuse boundaries. If the real interface is compositionally graded, a sharp step reference can artificially assign ordinary partitioning across the interface to Gibbsian excess. This problem has been highlighted more broadly in APT interface analysis, where the selected interface width and reference state strongly influence the quantitative interpretation of segregation \cite{Jenkins2020ReflectionsTomography}.

To avoid this artefact, the reference profile was instead described using a continuous composition-transition function, following the Cahn-style treatment of diffuse interfaces and the integral approach to interfacial excess described by Blum et al. \cite{Cahn1979,Huang2016TheCalculations,Blum2020AnSegregation}. In the present analysis, the transition function was constructed from the measured decomposed nitrogen profile:

\begin{equation}
X_N(z) =
\frac{
C_N^{\mathit{decomp}}(z)-C_N^{m}
}{
C_N^{n}-C_N^{m}
},
\label{eq:nitrogen_transition_function}
\end{equation}
where \(C_N^{\mathrm{decomp}}(z)\) is the total nitrogen profile after decomposition of nitrogen-containing molecular ions, and \(C_N^{m}\) and \(C_N^{n}\) are the far-field nitrogen concentrations in the matrix and nitride, respectively. In the numerical implementation, \(X_N(z)\) was bounded between 0 and 1 so that the continuous reference interpolated between, rather than extrapolated beyond, the measured matrix and nitride plateau compositions. The function \(X_N(z)\) was used only to describe the chemical transition and was not used to determine the physical nitride fraction, volume or atomic number density. The continuous reference profile for each element was then defined as
\begin{equation}
C_i^{\mathrm{ref}}(z)
=
C_i^{m}
+
\left(
C_i^{n}-C_i^{m}
\right)
X_N(z),
\label{eq:continuous_reference_profile}
\end{equation}
where \(C_i^{m}\) and \(C_i^{n}\) are the far-field concentrations of element \(i\) in the matrix and nitride, respectively. This reference describes the concentration profile expected from diffuse partitioning across the matrix--nitride interface in the absence of additional interfacial segregation. Concentrations were expressed as atomic fractions in the excess calculation. An error-function fit to monotonic Ti and Nb partitioning profiles was used only to estimate the interface centre, \(z_0\), and characteristic width, \(\lambda\), and was not used as the final reference profile in the excess integral. The displayed interface band was used only for visualisation and was not used as an integration limit.

A further correction was required because the nitride and matrix have different atomic number densities. Corrections to APT reconstructions using simulation and atomic-density information have previously been demonstrated \cite{Vurpillot2004ImprovementProbe,Grenier20143DTomography}. In Eq.~\ref{eq:gibbs_step}, the density term is treated as a constant. For a metal/nitride interface, this assumption is not valid because the number of atoms per unit volume changes across the interface. For the cubic phases considered here, the atomic number density was estimated from the measured lattice parameter as
\begin{equation}
n =
\frac{N_{\mathrm{cell}}}{a^3},
\label{eq:number_density_from_lattice}
\end{equation}
where \(N_{\mathrm{cell}}\) is the number of atoms per unit cell and \(a\) is the cubic lattice parameter. For the fcc matrix, \(N_{\mathrm{cell}}=4\), while for the B1 nitride, \(N_{\mathrm{cell}}=8\), assuming full occupancy of four metal atoms and four nitrogen atoms per rock-salt unit cell.

The corresponding matrix and nitride atomic number densities are denoted by \(n_m\) and \(n_n\), respectively. Using the 4D-STEM-measured lattice parameters, \(a_{\gamma}=\SI{3.70}{\angstrom}\) and \(a_{\mathrm{B1}}=\textcolor{black}{4.23\pm0.08~\si{\angstrom}}\), the calculated values were \(\textcolor{black}{n_m=79\pm3}~\mathrm{atoms\,nm^{-3}}\) and \(n_n=\textcolor{black}{106\pm6}~\mathrm{atoms\,nm^{-3}}\), respectively.

Because N is under-counted, the physical nitride fraction was determined from Ti+Nb relative to all consistently ranged alloy cations:
\begin{equation}
\textcolor{black}{
R(z)=
\frac{N_{\mathrm{Ti}}(z)+N_{\mathrm{Nb}}(z)}
{N_c(z)},
}
\label{eq:cation_marker}
\end{equation}
\textcolor{black}{where \(N_c(z)\) is the summed count of all consistently ranged cation alloy species, and \(R_m\) and \(R_n\) are the corresponding far-field matrix and nitride ratios. The matrix and B1 cation-site densities were \(n_{c,m}=4/a_{\gamma}^{3}=79.0~\mathrm{cations\,nm^{-3}}\) and \(n_{c,n}=4/a_{\mathrm{B1}}^{3}=52.9\pm3.0~\mathrm{cations\,nm^{-3}}\), respectively. The local B1 volume fraction was then}
\begin{equation}
\textcolor{black}{
\phi(z)=
\frac{
n_{c,m}\left[R(z)-R_m\right]
}{
n_{c,m}\left[R(z)-R_m\right]
+n_{c,n}\left[R_n-R(z)\right]
},
}
\label{eq:nitride_fraction}
\end{equation}
with \(\phi(z)\) bounded between 0 and 1. The local atomic number density was therefore
\begin{equation}
\textcolor{black}{
n(z)=
\left[1-\phi(z)\right]n_m
+\phi(z)n_n.
}
\label{eq:density_profile}
\end{equation}
The calculation assumes no substantial preferential loss among the ranged cations.

The same cation balance was applied to the precipitate to obtain its physical volume and equivalent-sphere diameter:
\begin{equation}
\textcolor{black}{
\begin{aligned}
N_{c,n} & = N_{c,\mathrm{ROI}}
\frac{R_{\mathrm{ROI}}-R_m}{R_n-R_m},\\
V_n & = \frac{N_{c,n}}{\eta n_{c,n}},\\
d_{\mathrm{eq}} & = \left(\frac{6V_n}{\pi}\right)^{1/3},
\end{aligned}
}
\label{eq:precipitate_size}
\end{equation}
where \(\eta=0.52\) is the detector efficiency. This gave \(V_n=(5.21\pm0.30)\times10^{3}~\mathrm{nm^3}\) and \(d_{\mathrm{eq}}=21.5\pm0.4~\si{\nano\metre}\). The quoted uncertainty was propagated from the measured B1 lattice parameter, with \(\eta\) and the cation assignment held fixed.

The final density-corrected interfacial excess was therefore calculated as
\begin{equation}
\Gamma_i =
\int_{z_{\mathrm{min}}}^{z_{\mathrm{max}}}
n(z)
\left[
C_i(z)-C_i^{\mathrm{ref}}(z)
\right]
\,\mathrm{d}z.
\tag{\ref{eq:gibbs_density_corrected}}
\end{equation}
This expression subtracts the continuous partitioning background before integrating the residual concentration, while also accounting for the density difference between the adjoining phases. It therefore gives a more conservative measure of true interfacial segregation than a conventional step-function analysis. This distinction is important beyond the present alloy, because metal/nitride interfaces in oxidised or nitrided alloys often combine diffuse chemistry, density changes and reconstruction artefacts, making conventional step-function excess values prone to overestimating or misassigning segregation.

\subsection{AI assistance}
{\emergencystretch=2em \tolerance=1700 ChatGPT-6 (OpenAI) was used to assist with manu\-script formatting and assist with code debugging. The authors reviewed and verified all incorporated suggestions and take responsibility for the final manuscript.\par}

\section*{Data availability}
Underlying data are available on request from Cameron Crabb
(c.crabb23@imperial.ac.uk).

\section*{Code availability}
The code used for data analysis is provided in the supplementary
files accompanying this manuscript and is also available from
Cameron Crabb (c.crabb23@imperial.ac.uk) upon request.
\medskip

% Compact bibliography typography; main text retains its original size.
\renewcommand{\bibfont}{\fontsize{8}{9.2}\selectfont}
\setlength{\bibsep}{0pt}
\bibliographystyle{elsarticle-num-condensed}
\bibliography{references}
\balance

\section*{Acknowledgements}
This study was supported by Rolls-Royce plc., The Royal Society (Tata University Research Fellowship URF/\allowbreak{}R1/\allowbreak{}201318 and Royal Society Enhancement Award RF/\allowbreak{}ERE/\allowbreak{}210200EM1), ERC (CoG DISCO grant 101171966) and EPSRC EP/\allowbreak{}V007661/\allowbreak{}1.

\section*{Author contributions}
C.C., M.S.C. and D.D. conceived the study.
C.C., G.T. and M.S.C. developed the S/TEM techniques and
correlative workflow. J.O.D. and C.C. developed the APT
analysis methodology. D.D., M.C.H. and C.C. developed the
underlying alloy, processing and testing methodologies.
All authors contributed to the writing and editing of
the manuscript.
\section*{Competing interests}
The authors declare no competing interests.

\end{document}